\documentclass[12pt]{article}

\usepackage{axodraw}
\usepackage{graphicx}

\makeatletter

\def\paragraph{\@startsection{paragraph}{4}{\z@}{+2.00ex plus
 +1ex minus +.2ex}{1.5ex plus .2ex}{\it\normalsize}}

\def\section{\@startsection {section}{1}{\z@}{+3.0ex plus +1ex minus
  +.2ex}{2.3ex plus .2ex}{\normalsize\bf\boldmath}}
\def\subsection{\@startsection{subsection}{2}{\z@}{+2.5ex plus +1ex
minus +.2ex}{1.5ex plus .2ex}{\normalsize\bf\boldmath}}
\def\subsubsection{\@startsection{subsubsection}{3}{\z@}{+3.25ex plus
 +1ex minus +.2ex}{1.5ex plus .2ex}{\normalsize\it}}

\expandafter\ifx\csname mathrm\endcsname\relax\def\mathrm#1{{\rm #1}}\fi

\newcounter{saveeqn}

\@addtoreset{equation}{section}

\newcount\@tempcntc
\def\@citex[#1]#2{\if@filesw\immediate\write\@auxout{\string\citation{#2}}\fi
  \@tempcnta\z@\@tempcntb\m@ne\def\@citea{}\@cite{\@for\@citeb:=#2\do
    {\@ifundefined
       {b@\@citeb}{\@citeo\@tempcntb\m@ne\@citea
        \def\@citea{,\penalty\@m\ }{\bf ?}\@warning
       {Citation `\@citeb' on page \thepage \space undefined}}%
    {\setbox\z@\hbox{\global\@tempcntc0\csname
b@\@citeb\endcsname\relax}%
     \ifnum\@tempcntc=\z@ \@citeo\@tempcntb\m@ne
       \@citea\def\@citea{,\penalty\@m}
       \hbox{\csname b@\@citeb\endcsname}%
     \else
      \advance\@tempcntb\@ne
      \ifnum\@tempcntb=\@tempcntc
      \else\advance\@tempcntb\m@ne\@citeo
      \@tempcnta\@tempcntc\@tempcntb\@tempcntc\fi\fi}}\@citeo}{#1}}

\def\@citeo{\ifnum\@tempcnta>\@tempcntb\else\@citea
  \def\@citea{,\penalty\@m}%
  \ifnum\@tempcnta=\@tempcntb\the\@tempcnta\else
   {\advance\@tempcnta\@ne\ifnum\@tempcnta=\@tempcntb \else
\def\@citea{--}\fi
    \advance\@tempcnta\m@ne\the\@tempcnta\@citea\the\@tempcntb}\fi\fi}

\newcommand{\gsim}
{\mathrel{\raisebox{-.3em}{$\stackrel{\displaystyle >}{\sim}$}}}
\def\asymp#1%
{\mathrel{\raisebox{-.4em}{$\widetilde{\scriptstyle #1}$}}}

\def\Nequal#1%
{\mathrel{\raisebox{-.5em}{$\stackrel{=}{\scriptstyle\rm#1}$}}}
\newcommand{\dsl}[1]{\not \hspace{-0.7mm}#1}
\def\dsl{\mathpalette\make@slash}
\def\make@slash#1#2{\setbox\z@\hbox{$#1#2$}%
  \hbox to 0pt{\hss$#1/$\hss\kern-\wd0}\box0}

\def\beq{\begin{equation}}
\def\eeq{\end{equation}}
\def\beqar{\begin{eqnarray}}
\def\eeqar{\end{eqnarray}}
\def\barr#1{\begin{array}{#1}}
\def\earr{\end{array}}
\def\bfi{\begin{figure}}
\def\efi{\end{figure}}
\def\btab{\begin{table}}
\def\etab{\end{table}}
\def\bce{\begin{center}}
\def\ece{\end{center}}

\def\text{\textstyle}

\def\de{\delta}

\def\si{\sigma}

\def\refeq#1{\mbox{(\ref{#1})}}

\def\reffis#1{\mbox{Figures~\ref{#1}}}
\def\refta#1{\mbox{Table~\ref{#1}}}

\def\citere#1{\mbox{Ref.~\cite{#1}}}
\def\citeres#1{\mbox{Refs.~\cite{#1}}}

\newcommand{\TeV}{\unskip\,\mathrm{TeV}}
\newcommand{\GeV}{\unskip\,\mathrm{GeV}}
\newcommand{\MeV}{\unskip\,\mathrm{MeV}}

\newcommand{\fb}{\unskip\,\mathrm{fb}}

\newcommand{\Oa}{\mathswitch{{\cal{O}}(\alpha)}}

\renewcommand{\L}{{\cal{L}}}

\def\mathswitchr#1{\relax\ifmmode{\mathrm{#1}}\else$\mathrm{#1}$\fi}

\newcommand{\PW}{\mathswitchr W}
\newcommand{\Pw}{\mathswitchr w}
\newcommand{\PZ}{\mathswitchr Z}

\newcommand{\PH}{\mathswitchr H}
\newcommand{\Pe}{\mathswitchr e}

\newcommand{\Pd}{\mathswitchr d}

\newcommand{\Pu}{\mathswitchr u}

\newcommand{\Ps}{\mathswitchr s}

\newcommand{\Pc}{\mathswitchr c}

\newcommand{\Pb}{\mathswitchr b}

\newcommand{\Pt}{\mathswitchr t}

\newcommand{\Pep}{\mathswitchr {e^+}}
\newcommand{\Pem}{\mathswitchr {e^-}}

\def\mathswitch#1{\relax\ifmmode#1\else$#1$\fi}

\newcommand{\MW}{\mathswitch {M_\PW}}

\newcommand{\MZ}{\mathswitch {M_\PZ}}
\newcommand{\MH}{\mathswitch {M_\PH}}
\newcommand{\Me}{\mathswitch {m_\Pe}}

\newcommand{\Md}{\mathswitch {m_\Pd}}
\newcommand{\Mu}{\mathswitch {m_\Pu}}
\newcommand{\Ms}{\mathswitch {m_\Ps}}
\newcommand{\Mc}{\mathswitch {m_\Pc}}
\newcommand{\Mb}{\mathswitch {m_\Pb}}
\newcommand{\Mt}{\mathswitch {m_\Pt}}

\newcommand{\sw}{\mathswitch {s_\Pw}}

\newcommand{\GF}{\mathswitch {G_\mu}}

\newcommand{\alphas}{\alpha_{\mathrm{s}}}

\def\solid{\raise.9mm\hbox{\protect\rule{1.1cm}{.2mm}}}
\def\dash{\raise.9mm\hbox{\protect\rule{2mm}{.2mm}}\hspace*{1mm}}

\def\ie{i.e.\ }

\def\cf{cf.\ }

\newcommand{\FSR}{{\mathrm{FSR}}}
\newcommand{\QCD}{{\mathrm{QCD}}}

\newcommand{\eennh}{\Pep\Pem\to\nu\bar\nu\PH}

\newcommand{\eetth}{\Pep\Pem\to\Pt\bar\Pt\PH}

\def\draftdate{\relax}
\def\mda{\relax}
\def\mua{\relax}
\def\mla{\relax}
\def\draft{
\def\thtystars{******************************}
\def\sixtystars{\thtystars\thtystars}
\typeout{}
\typeout{\sixtystars**}
\typeout{* Draft mode!
         For final version remove \protect\draft\space in source file *}
\typeout{\sixtystars**}
\typeout{}
\def\draftdate{\today}
\def\mua{\marginpar[\boldmath\hfil$\uparrow$]%
                   {\boldmath$\uparrow$\hfil}%
                    \typeout{marginpar: $\uparrow$}\ignorespaces}
\def\mda{\marginpar[\boldmath\hfil$\downarrow$]%
                   {\boldmath$\downarrow$\hfil}%
                    \typeout{marginpar: $\downarrow$}\ignorespaces}
\def\mla{\marginpar[\boldmath\hfil$\rightarrow$]%
                   {\boldmath$\leftarrow $\hfil}%
                    \typeout{marginpar: $\leftrightarrow$}\ignorespaces}
\def\Mua{\marginpar[\boldmath\hfil$\Uparrow$]%
                   {\boldmath$\Uparrow$\hfil}%
                    \typeout{marginpar: $\uparrow$}\ignorespaces}
\def\Mda{\marginpar[\boldmath\hfil$\Downarrow$]%
                   {\boldmath$\Downarrow$\hfil}%
                    \typeout{marginpar: $\downarrow$}\ignorespaces}
\def\Mla{\marginpar[\boldmath\hfil$\Rightarrow$]%
                   {\boldmath$\Leftarrow $\hfil}%
                    \typeout{marginpar: $\leftrightarrow$}\ignorespaces}
\overfullrule 5pt
\oddsidemargin -15mm
\marginparwidth 29mm
}

\def\stars{\strut\leaders\hbox{*}\hfill\strut}
\def\starline{\hfil\strut\hfil\hbox to \textwidth {\stars}\hfil}

\begin{document}
\thispagestyle{empty}
\def\thefootnote{\fnsymbol{footnote}}
\setcounter{footnote}{1}
\null
\draftdate\hfill KA-TP-05-2003\\
\strut\hfill MPP-2003-27 \\
\strut\hfill PSI-PR-03-12\\
%\strut\hfill LC-TH-2003-???\\
\strut\hfill hep-ph/0307193
\vfill
\begin{center}
{\Large \bf\boldmath
Electroweak radiative corrections to 
$\Pep\Pem\to \Pt\bar\Pt\PH$%
\par} 
\vspace{1cm}
{\large
{\sc A.\ Denner$^1$, S.\ Dittmaier$^2$, M. Roth$^3$ and 
M.M.~Weber$^1$} } \\[1cm]
$^1$ {\it Paul Scherrer Institut, W\"urenlingen und Villigen\\
CH-5232 Villigen PSI, Switzerland} \\[0.5cm]
$^2$ {\it Max-Planck-Institut f\"ur Physik 
(Werner-Heisenberg-Institut) \\
%F\"ohringer Ring 6, 
D-80805 M\"unchen, Germany}
\\[0.5cm]
$^3$ {\it Institut f\"ur Theoretische Physik, Universit\"at Karlsruhe \\
D-76128 Karlsruhe, Germany}
\par \vskip 1em
\end{center}\par
\vskip 2cm {\bf Abstract:} \par We have calculated the complete
electroweak ${\cal O}(\alpha)$ radiative corrections to the
Higgs-boson production process $\Pep\Pem\to\Pt\bar\Pt\PH$ in the
electroweak Standard Model. Initial-state radiation beyond $\Oa$ is
included in the structure-function approach.  The calculation of the
corrections is briefly described, and numerical results are presented
for the total cross section. Both the photonic and the genuine weak
corrections reach the order of about 10\% or even more and show a
non-trivial dependence on the Higgs-boson mass and on the scattering
energy.  We compare our results with two previous calculations that
obtained differing results at high energies.
\par
\vskip 2cm
\noindent
July 2003
\null
\setcounter{page}{0}
\clearpage
\def\thefootnote{\arabic{footnote}}
\setcounter{footnote}{0}

\section{Introduction}
\label{se:intro}

In the electroweak Standard Model (SM) all fermions $f$ receive their
masses $m_f$ via Yukawa couplings to the Higgs field.  Splitting the
Higgs field into its vacuum-expectation value
$v=(\sqrt{2}\GF)^{-1/2}\approx246\GeV$ and the physical excitation
$H(x)$, the Yukawa term in the SM Lagrangian reads
$\L_{\mathrm{Yuk}}=-\sum_f m_f(1+H(x)/v)\overline{\psi}_f\psi_f$.
Thus, the Yukawa coupling strength is predicted to be $m_f/v$ at tree
level, and the experimental determination of the Higgs--fermion
couplings represents an important test of the mass generation via the
Higgs mechanism.  Since the top quark is the heaviest of all fermions,
it is supposed to play a key role in a theory of fermion masses.
Therefore, the measurement of the top-quark Yukawa coupling is of
particular interest.  For not too large Higgs-boson masses, $\MH\sim
100$--$200\GeV$, a promising process for this task is $\eetth$, as
already pointed out in \citere{Gunion:1996vv}.  To achieve a
measurement with a precision of the order of $\sim 5\%$, however, an
$\Pep\Pem$ linear collider (LC) with high centre-of-mass (CM) energies
($\sqrt{s}\sim 800$--$1000\GeV$) and high luminosity ($L\sim
1000\fb^{-1}$) is required \cite{Baer:1999ge}.  A better accuracy
might be obtained by a simultaneous fit of various Higgs-boson
parameters to the whole profile of the Higgs boson at a LC
\cite{Battaglia:2000jb}.  Moreover, an investigation of the process
$\eetth$ might be useful for setting bounds on non-standard physics
\cite{Gunion:1996vv,Han:1999xd} in the top-quark--Higgs-boson
coupling.

A determination of the top-quark Yukawa coupling at the level of a few
per cent requires both a proper understanding of the background
\cite{Moretti:1999kx} to the decaying $\Pt\bar\Pt\PH$ final state and
a theoretical prediction of the $\eetth$ signal cross section within
per-cent accuracy.  Thus, radiative corrections have to be controlled
within this accuracy, a task that is rather complicated for a process
with three massive unstable particles in the final state.  As a first
step, the process $\eetth$ can be treated in the approximation of
stable top quarks and Higgs bosons in the final state.  The
corresponding lowest-order predictions are already known for a long
time \cite{Gaemers:1978jr}. These results were supplemented by the
${\cal O}(\alphas)$ QCD corrections to the total production cross
section in the SM, first within the ``effective Higgs-boson
approximation'' \cite{Dawson:1997im} that is valid only for small
Higgs-boson masses and very high energies,
and subsequently \cite{Dittmaier:1998dz,Dawson:1998ej}%
\footnote{In \citere{Dawson:1998ej} only the QCD corrections to the
  photon-exchange channel for $\Pt\bar\Pt\PH$ production were taken
  into account.}  based on the full set of QCD diagrams in ${\cal
  O}(\alphas)$.  The QCD corrections to Higgs-boson production in
association with heavy-quark pairs ($\Pt\bar\Pt/\Pb\bar\Pb$) in the
minimal supersymmetric SM were
discussed in \citeres{Dawson:1998qq,Dittmaier:2000tc,Zhu:2002iy}.%
\footnote{In \citere{Dawson:1998qq} only the photon-exchange channel
  is corrected, so that the relative correction given there coincides
  with the one obtained in \citere{Dawson:1998ej} where the SM process
  is treated analogously.  The calculation of
  \citere{Dittmaier:2000tc} includes the full set of ${\cal
    O}(\alphas)$ QCD diagrams, and in \citere{Zhu:2002iy} the SUSY-QCD
  corrections to $\eetth$ are considered.}  Besides corrections to
total cross sections, the QCD corrections to the Higgs-boson energy
distribution were discussed in \citere{Dittmaier:mg}, both for the SM
and its minimal supersymmetric extension.  Recently first results for
the electroweak $\Oa$ corrections to $\eetth$ in the SM have been
presented in \citeres{You:2003zq} and \cite{Belanger:2003nm}.  These
calculations were found to agree for small energies but to differ at
high energies.

In this paper we present results of a further, completely independent
calculation of the $\Oa$ electroweak corrections, which is
additionally improved by the leading higher-order corrections from
initial-state radiation. Details on this calculation, which we have
also implemented in a Monte Carlo event generator, will be given
elsewhere. Here we sketch only the main ingredients.  Moreover, we
compare our results with those of \citeres{You:2003zq} and
\cite{Belanger:2003nm}.  While we find good agreement with the results
of the recent calculation of \citere{Belanger:2003nm} at all
considered energies, our results differ from those of
\citere{You:2003zq} at high energies and close to threshold.
Moreover, we could reproduce the QCD corrections of
\citere{Dittmaier:1998dz} from our results on photonic final-state
radiation within statistical integration errors.

\section{Method of calculation}
\label{se:calc}

We have calculated the complete $\Oa$ electroweak virtual and real
photonic corrections to the process $\eetth$, following the same
strategy as used in our previous calculation of the corrections to
$\eennh$ \cite{Denner:2003yg}.

The calculation of the one-loop diagrams has been performed in the
't~Hooft--Feynman gauge both in the conventional and in the
background-field formalism using the conventions of
\citeres{Denner:1993kt} and \cite{Denner:1994xt}, respectively.  The
renormalization is carried out in the on-shell renormalization scheme,
as described there.  The electron mass $\Me$ is neglected whenever
possible.

The calculation of the Feynman diagrams has been performed in two
completely independent ways, leading to two independent computer codes
for the numerical evaluation. Both calculations are based on the
methods described in \citere{Denner:1993kt}.  Apart from the 5-point
functions the tensor coefficients of the one-loop integrals are
recursively reduced to scalar integrals with the Passarino--Veltman
algorithm \cite{Passarino:1979jh} at the numerical level.  The scalar
integrals are evaluated using the methods and results of
\citeres{Denner:1993kt,'tHooft:1979xw}, where ultraviolet divergences
are regulated dimensionally and IR divergences with an infinitesimal
photon mass. The 5-point functions are reduced to 4-point functions
following \citere{Denner:2002ii}, where a method for a direct
reduction is described that avoids leading inverse Gram determinants
which potentially cause numerical instabilities.  The two calculations
differ in the following points.  In the first calculation, the Feynman
graphs are generated with {\sl Feyn\-Arts} version 1.0
\cite{Kublbeck:1990xc}.  Using {\sl Mathematica} the amplitudes are
expressed in terms of standard matrix elements and coefficients of
tensor integrals.  The whole calculation has been carried out in the
conventional and in the background-field formalism.  The second
calculation has been done with the help of {\sl Feyn\-Arts} version 3
\cite{Hahn:2000kx}, and the analytical expressions have been generated
by {\sl FormCalc} \cite{Hahn:1998yk} and translated into a {\sl C}
code. The scalar and tensor coefficients, in particular those for the
5-point functions, have been evaluated by own routines, as described
above.

The results of the two different codes, and also those obtained within
the conventional and background-field formalism, are in good numerical
agreement (typically within at least 12 digits for non-exceptional
phase-space points and double precision).  The agreement of the
results in the conventional and background-field formalism, in
particular, checks the gauge independence of our results.

The matrix elements for the real photonic corrections are evaluated
using the Weyl--van der Waerden spinor technique as formulated in
\citere{Dittmaier:1999nn} and have been successfully checked against
the result obtained with the package {\sl Madgraph}
\cite{Stelzer:1994ta}.  The soft and collinear singularities are
treated %both 
in the dipole subtraction method \cite{Dittmaier:2000mb}.
%and in the phase-space slicing method following closely \citere{bo93}.
Beyond $\Oa$ initial-state-radiation (ISR) corrections are included at
the leading-logarithmic level using the structure functions given in
\citere{lep2repWcs} (for the original papers see references therein).

The phase-space integration is performed with Monte Carlo techniques
in both computer codes. The first code employs a multi-channel Monte
Carlo generator similar to the one implemented in {\sl RacoonWW}
\cite{Roth:1999kk,Denner:1999gp} and {\sl Lusifer}
\cite{Dittmaier:2002ap}, the second one uses the adaptive
multi-dimensional integration program {\sl VEGAS}
\cite{Lepage:1977sw}.

\section{Numerical results}
\label{se:numres}

For the numerical evaluation we use the following set of
SM parameters \cite{Hagiwara:pw},
\beq
\begin{array}[b]{lcllcllcl}
\GF & = & 1.16639 \times 10^{-5} \GeV^{-2}, \quad&
\alpha(0) &=& 1/137.03599976, \quad&
\alphas(\MZ) &=& 0.1172, \\
\MW & = & 80.423\GeV, &
\MZ & = & 91.1876\GeV, & & &  \\
\Me & = & 0.510998902\MeV, &
m_\mu &=& 105.658357\MeV,\quad &
m_\tau &=& 1.77699\GeV, \\
\Mu & = & 66\MeV, &
\Mc & = & 1.2\GeV, &
\Mt & = & 174.3\;\GeV, \\
\Md & = & 66\MeV, &
\Ms & = & 150\MeV, &
\Mb & = & 4.3\GeV. 
\end{array}
\label{eq:SMpar}
\eeq
We do not calculate the W-boson mass from $\GF$ but use its
experimental value as input. The masses of the light quarks are
adjusted to reproduce the hadronic contribution to the photonic vacuum
polarization of \citere{Jegerlehner:2001ca}.  We parametrize the
lowest-order cross section with the Fermi constant $\GF$
($\GF$-scheme), \ie we derive the electromagnetic coupling $\alpha$
according to $ \alpha_{\GF} = \sqrt{2}\GF\MW^2\sw^2/\pi$.  This, in
particular, absorbs the running of the electromagnetic coupling
$\alpha(Q^2)$ from $Q=0$ to the electroweak scale ($Q\sim\MZ$) into
the lowest-order cross section so that the results are practically
independent of the masses of the light quarks.  In the relative
radiative corrections, we use, however, $\alpha(0)$ as coupling
parameter, which is the correct effective coupling for real photon
emission.

In the following we separate the {\it photonic} corrections, which
comprise loop diagrams with virtual photon exchange in the loop and
the corresponding parts of the counter terms as well as real photon
emission, from the full electroweak ${\cal O}(\alpha)$ corrections;
the remaining non-photonic electroweak corrections are called {\it
  weak}.  Since the lowest-order diagrams involve only neutral-current
couplings, but no W-boson exchange, this splitting is gauge invariant.
Moreover, we separately discuss the {\it higher-order\/} [i.e.\ beyond
${\cal O}(\alpha)$] {\it ISR} corrections that are obtained from the
convolution of the lowest-order cross section with the
leading-logarithmic structure functions, but with the ${\cal
  O}(\alpha)$ contribution subtracted.  The higher-order ISR is
included in the electroweak corrections shown in the following plots.

In addition to the electroweak corrections, we also include results on
the QCD corrections which can be deduced from the part of the photonic
corrections that is proportional to $Q_\Pt^2$, where $Q_\Pt=2/3$ is
the relative electric charge of the top quark, \ie from the
final-state radiation (FSR).  The QCD correction is obtained from
these corrections upon replacing the factor $Q_\Pt^2\alpha$ by
$C_{\mathrm{F}}\alphas=4\alphas(\mu^2)/3$.
%The QCD corrections which are obtained in this way have been compared
%with the (independent) results of \citere{Dittmaier:1998dz}, and numerical
%agreement within the integration errors was found. 
Following \citere{Dittmaier:1998dz}, the QCD renormalization scale
$\mu$ is set to the CM energy in the following, and the running of the
strong coupling is evaluated at the two-loop level
($\overline{\mathrm{MS}}$ scheme) with five active flavours,
normalized by $\alphas(\MZ^2)$ as given in Eq.~\refeq{eq:SMpar}.  For
$\sqrt{s}=500, 800$, and $1000\GeV$ the resulting values for the
strong coupling are given by $\alphas(\MZ^2)= 0.09349, 0.08857$, and
$0.08642$, respectively.

\begin{figure}
\centerline{\includegraphics[width=.5\textwidth,bb=78 415 279 628]{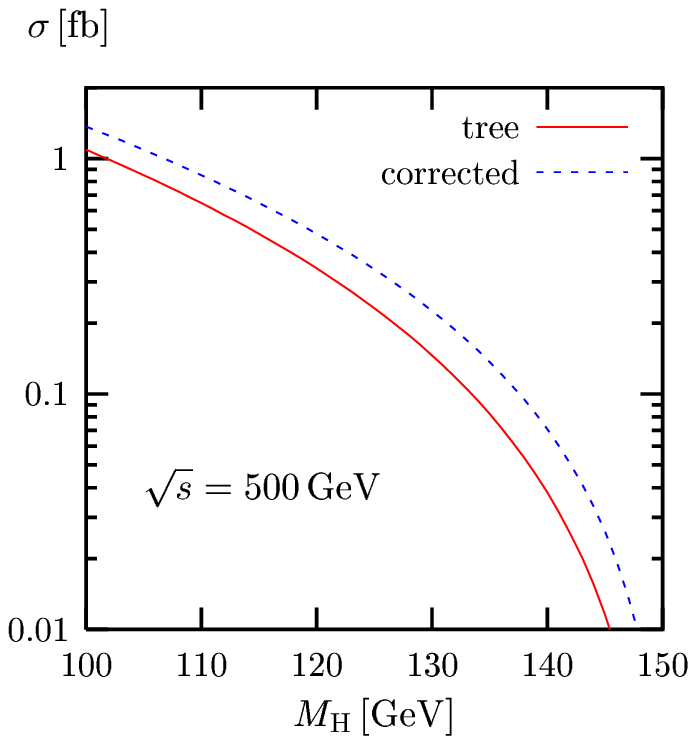}
            \includegraphics[width=.5\textwidth,bb=78 415 279 628]{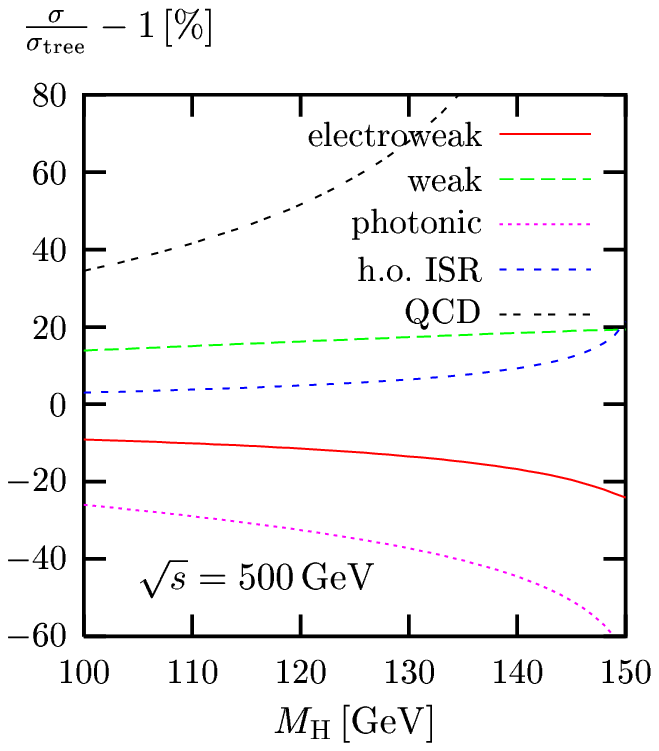}}
\caption{Lowest-order and corrected cross sections (l.h.s.) as well as
  relative corrections (r.h.s.) in the $\GF$-scheme for a CM energy
  $\sqrt{s}=500\GeV$}
\label{fi:cs500}
\end{figure}

\begin{figure}
\centerline{\includegraphics[width=.5\textwidth,bb=78 415 279 628]{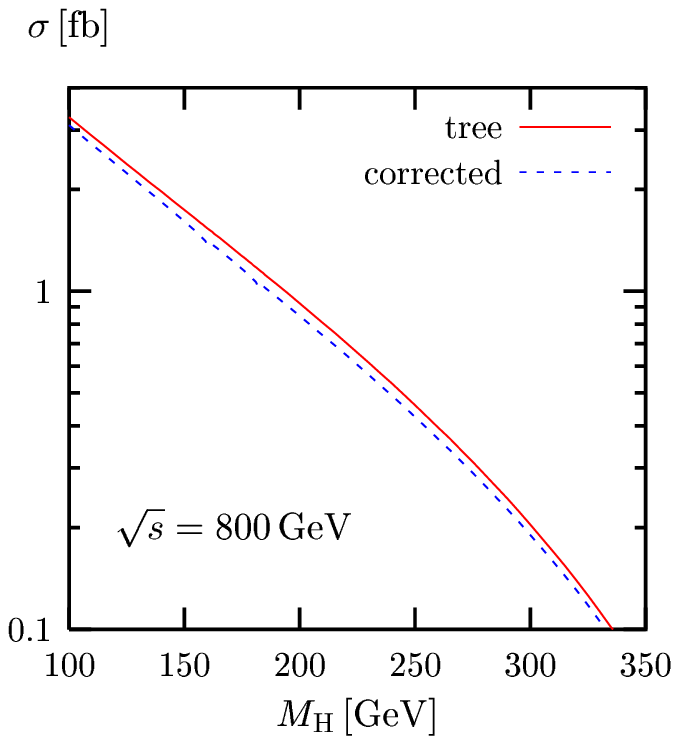}
            \includegraphics[width=.5\textwidth,bb=78 415 279 628]{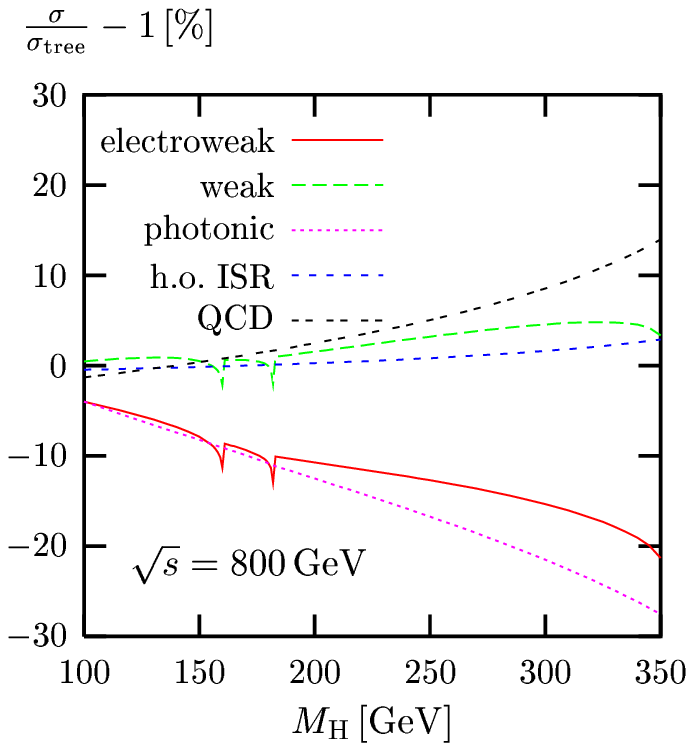}}
\caption{Lowest-order and corrected cross sections (l.h.s.) as well as
  relative corrections (r.h.s.) in the $\GF$-scheme for a CM energy
  $\sqrt{s}=800\GeV$}
\label{fi:cs800}
\end{figure}

\begin{figure}
\centerline{\includegraphics[width=.5\textwidth,bb=78 415 279 628]{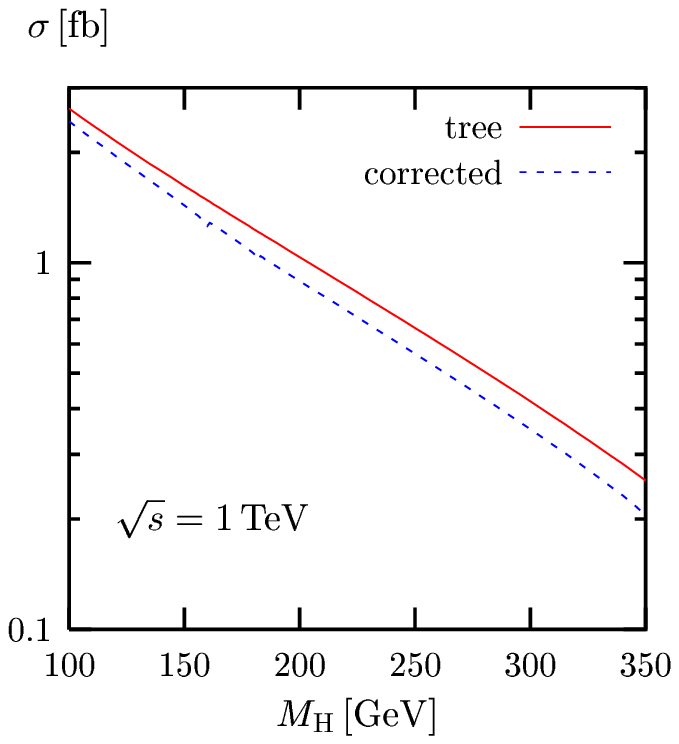}
            \includegraphics[width=.5\textwidth,bb=78 415 279 628]{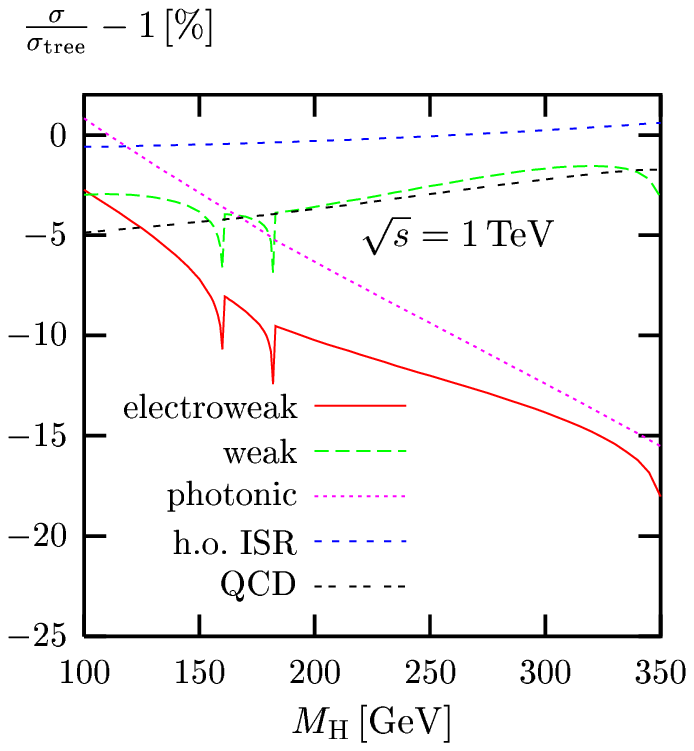}}
\caption{Lowest-order and corrected cross sections (l.h.s.) as well as
  relative corrections (r.h.s.) in the $\GF$-scheme for a CM energy
  $\sqrt{s}=1\TeV$}
\label{fi:cs1000}
\end{figure}

\begin{figure}
\centerline{\includegraphics[width=.5\textwidth,bb=78 415 279 628]{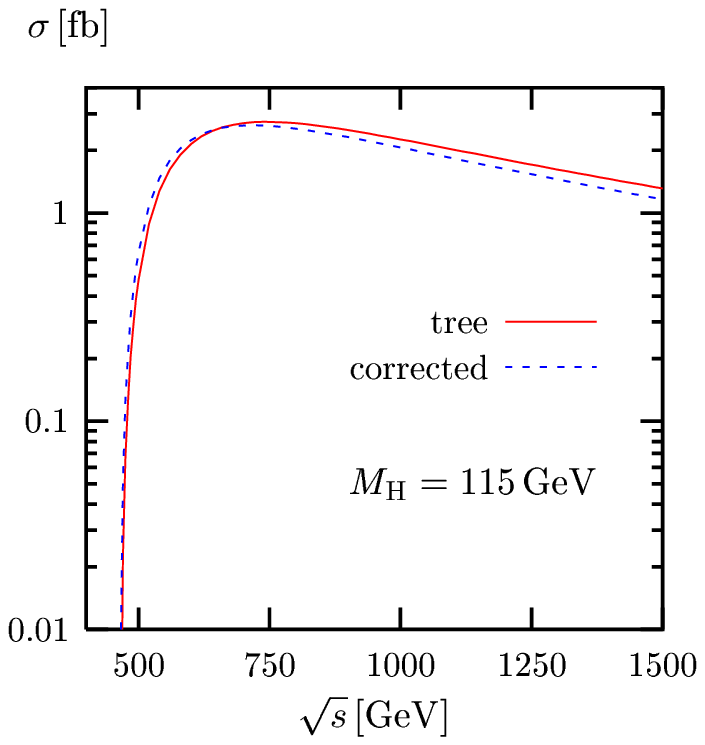}
            \includegraphics[width=.5\textwidth,bb=78 415 279 628]{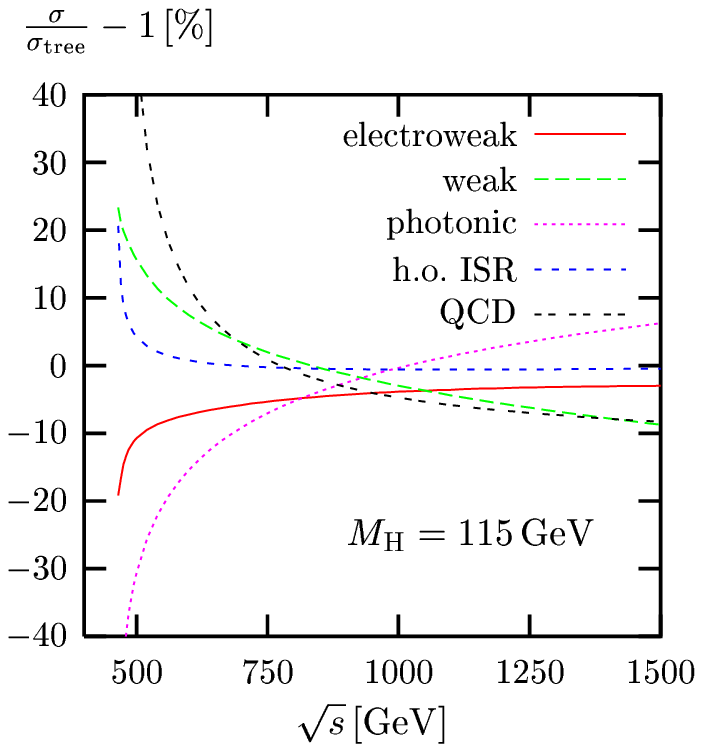}}
\caption{Lowest-order and corrected cross sections (l.h.s.) as well as
  relative corrections (r.h.s.) in the $\GF$-scheme for a Higgs-boson
  mass $\MH=115\GeV$}
\label{fi:mh115}
\end{figure}

\begin{figure}
\centerline{\includegraphics[width=.5\textwidth,bb=78 415 279 628]{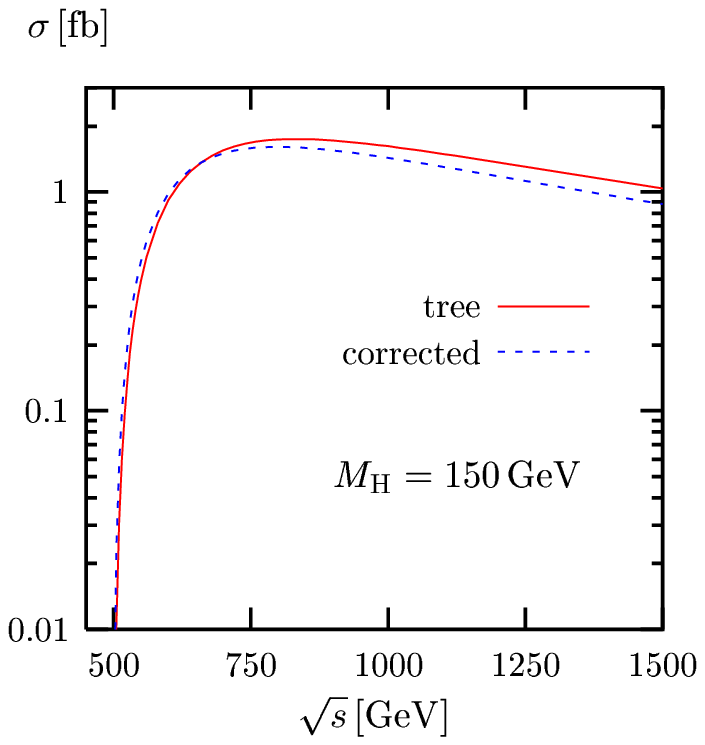}
            \includegraphics[width=.5\textwidth,bb=78 415 279 628]{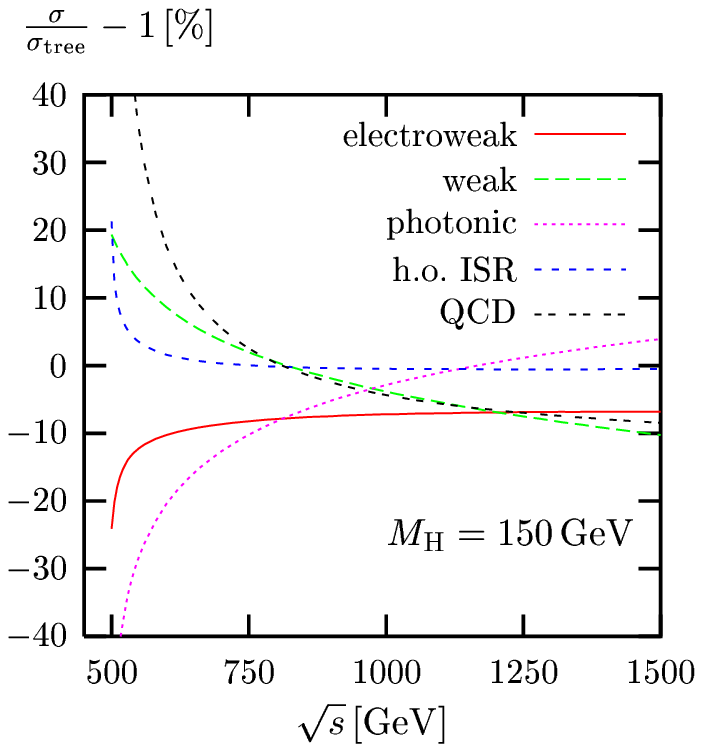}}
\caption{Lowest-order and corrected cross sections (l.h.s.) as well as
  relative corrections (r.h.s.) in the $\GF$-scheme for a Higgs-boson
  mass $\MH=150\GeV$}
\label{fi:mh150}
\end{figure}

\begin{figure}
\centerline{\includegraphics[width=.5\textwidth,bb=78 415 279 628]{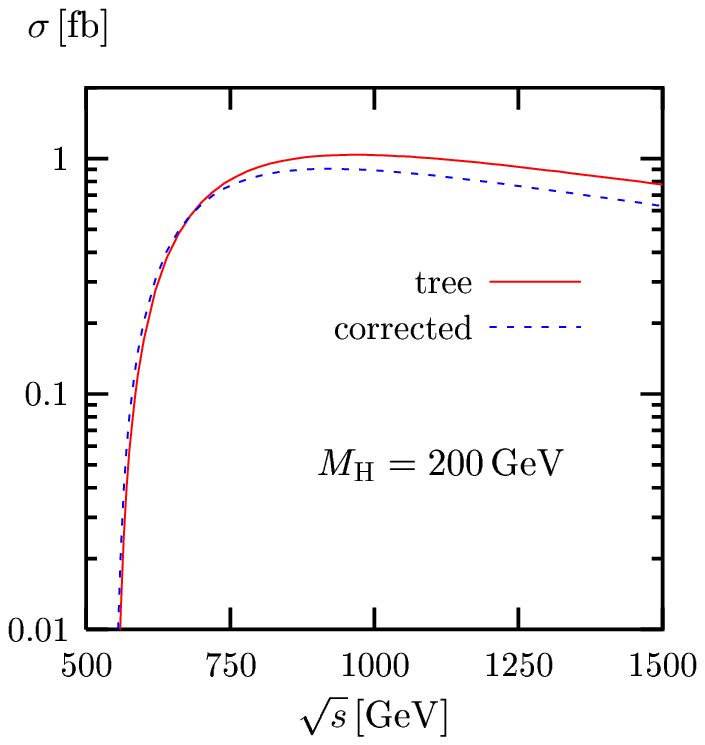}
            \includegraphics[width=.5\textwidth,bb=78 415 279 628]{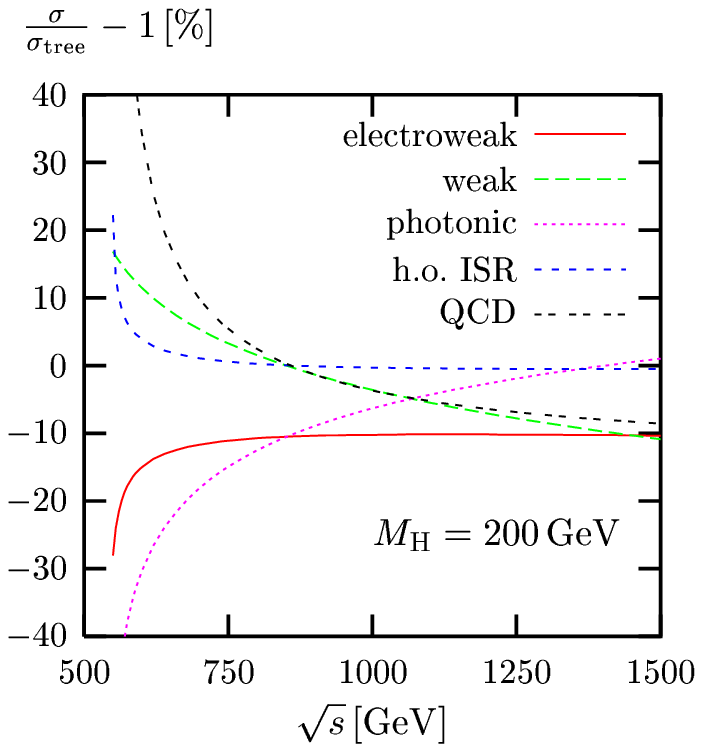}}
\caption{Lowest-order and corrected cross sections (l.h.s.) as well as
  relative corrections (r.h.s.) in the $\GF$-scheme for a Higgs-boson
  mass $\MH=200\GeV$}
\label{fi:mh200}
\end{figure}

In this letter, we consider merely total cross sections without any
cuts; distributions will be discussed elsewhere.  For reference we
give some numbers for the total cross section in lowest order,
$\si_\mathrm{tree}$, the cross section including electroweak and QCD
corrections, $\si_\mathrm{corr}$, and the various contributions to the
relative corrections defined as $\de=\sigma/\sigma_{\mathrm{tree}}-1$
in \refta{ta:xsection}.
\begin{table}
\newdimen\digitwidth
\setbox0=\hbox{0}
\digitwidth=\wd0
\catcode`!=\active
\def!{\kern\digitwidth}
\newdimen\minuswidth
\setbox0=\hbox{$-$}
\minuswidth=\wd0
\catcode`?=\active
\def?{\kern\minuswidth}
\begin{tabular}{rccc}
\hline
$\sqrt{s}$ [GeV] & \multicolumn{3}{c}{500} 
\\
$\MH$ [GeV] & 115 & 125 & 140 \\
\hline
$\sigma_{\mathrm{tree}}$ [fb] & 0.47901(7) & 0.23150(3) & 0.038189(6) 
\\
$\sigma_{\mathrm{corr}}$ [fb] & 0.6506(6)! & 0.3401(4)! & 0.0713(1)!! \\
\hline
$\de_{\mathrm{QCD}}[\%]$ & ?46.5(1)!! & ?59.2(2)!! & 
\kern-\digitwidth?103.4(3)!!
\\
$\de_{\mathrm{phot}}[\%]$ & $-30.64(1)!$ & $-34.69(1)!$ & $-44.51(1)!$
\\
$\de_{\mathrm{hoISR}}[\%]$ & ?!4.25(3)! & ?!5.53(3)! & ?!9.28(3)!
\\
$\de_{\mathrm{weak}}[\%]$  & ?15.70(3)! & ?16.85(3)! & ?18.51(3)!
\\
\hline
\end{tabular}
\\[2em]
\begin{tabular}{rcccccc}
\hline
$\sqrt{s}$ [GeV] & \multicolumn{6}{c}{800}
\\
$\MH$ [GeV] & 115 & 150 & 200 & 250 & 300 & 350 \\
\hline
$\sigma_{\mathrm{tree}}$ [fb] &  
2.7004(4)  &  1.7406(3)  & 0.9217(1)  & 0.46076(6)  & 0.20432(3)  & 0.07165(1)
\\
$\sigma_{\mathrm{corr}}$ [fb]  &               
2.541(1)!  &  1.6076(7)  & 0.8443(4)  & 0.4251(2)!  & 0.1904(1)!  & 0.06648(6)
\\
\hline
$\de_{\mathrm{QCD}}[\%]$ & 
$-0.87(2)$ & ?0.36(3) & ?2.43(4) & ?!5.03(4) & ?!8.58(6) & ?14.12(6)
\\
$\de_{\mathrm{phot}}[\%]$ & 
$-5.30(1)$ & $-8.26(1)$ & $-12.54(1)$ & $-16.76(1)$ & $-21.54(1)$ & $-27.55(1)$
\\
$\de_{\mathrm{hoISR}}[\%]$ & 
$-0.41(3)$ & $-0.19(3)$ & ?0.22(3) & ?!0.79(3) & ?!1.59(3) & ?!2.86(3)
\\
$\de_{\mathrm{weak}}[\%]$ & ?0.69(1) & ?0.45(1) & ?1.49(1) & ?!3.19(1) & ?!4.56(1) & ?!3.36(5)
\\
\hline
\end{tabular}
\\[2em]
\begin{tabular}{rcccccc}
\hline
$\sqrt{s}$ [GeV] & \multicolumn{6}{c}{1000}
\\
$\MH$ [GeV] & 115 & 150 & 200 & 250 & 300 & 350 \\
\hline
$\sigma_{\mathrm{tree}}$ [fb] & 
2.2594(3)    & 1.6208(2)    & 1.0356(2)    & 0.6643(1)    & 0.41894(6)   & 0.25510(4)
\\
$\sigma_{\mathrm{corr}}$ [fb] &                
2.061(1)!    & 1.4311(7)    & 0.8900(5)    & 0.5639(3)    & 0.3510(2)!   & 0.2045(2)!
\\
\hline
$\de_{\mathrm{QCD}}[\%]$ & 
$-4.76(2)$ & $-4.37(2)$ & $-3.72(3)$ & $-2.99(3)$ & $!-2.26(4)$ & $!-1.73(5)$
\\
$\de_{\mathrm{phot}}[\%]$ & 
$-0.34(1)$ & $-2.89(1)$ & $-6.35(1)$ & $-9.41(1)$ & $-12.43(1)$ & $-15.55(1)$
\\
$\de_{\mathrm{hoISR}}[\%]$ & 
$-0.61(3)$ & $-0.52(3)$ & $-0.34(3)$ & $-0.11(3)$ & $!?0.19(3)$ & ?!0.57(3)
\\
$\de_{\mathrm{weak}}[\%]$ & 
$-3.06(2)$ & $-3.91(2)$ & $-3.66(3)$ & $-2.61(2)$ & $!-1.71(2)$ & $!-3.13(3)$ 
\\
\hline
\end{tabular}
\caption{Lowest-order cross section for $\eetth$ in the $\GF$-scheme,
  $\si_{\mathrm{tree}}$, cross section including electroweak and QCD
  corrections, $\si_{\mathrm{corr}}$, and various contributions to the
  relative corrections $\de$ (as described in the text)
  for various Higgs-boson masses at $\sqrt{s}=500\GeV$, $800\GeV$, and $1\TeV$}
\label{ta:xsection}
\end{table}
The last numbers in parentheses correspond to the Monte Carlo integration
error of the last given digits.
In \reffis{fi:cs500}--\ref{fi:cs1000} we show the lowest-order cross
section
as well as the 
corresponding
corrections as a function of the Higgs-boson mass for
the typical LC CM energies $\sqrt{s}=500\GeV$, $800\GeV$, and $1\TeV$.
In \reffis{fi:mh115}--\ref{fi:mh200} the results are illustrated for
fixed Higgs-boson masses of $\MH=115\GeV,150\GeV$, and $200\GeV$.

Away from the kinematic threshold at $\sqrt{s}=2\Mt+\MH$ the total
cross section is typically of the order of 0.5--$3\fb$ and becomes
maximal in the energy range between $700\GeV$ and $1\TeV$ for small
Higgs-boson masses.  As already discussed in
\citeres{Dittmaier:1998dz,Dawson:1998ej} in detail, the QCD
corrections are positive and rather large in the threshold region
($\sqrt{s}\gsim 2\Mt+\MH$), where soft-gluon exchange between in the
$\Pt\bar\Pt$ system leads to a Coulomb-like singularity. Away from
threshold this singularity is diluted by the larger phase space, since
the singularity demands a low relative velocity of the quarks.
Averaging the singular factor over the phase space leads to the
following threshold behaviour of the QCD correction
\cite{Dittmaier:1998dz},
\beq
\delta_{\QCD} \;\sim\; \frac{32\alphas}{9\beta_\Pt}, \qquad
\delta_{\FSR} \;\sim\; \frac{8Q_\Pt^2\alpha}{3\beta_\Pt}, \qquad
\beta_\Pt = \frac{\sqrt{(\sqrt{s}-\MH)^2-4\Mt^2}}{2\Mt},
\eeq
where $\beta_\Pt$ is the maximal quark velocity in the $\Pt\bar\Pt$
rest frame. The FSR corrections show the same behaviour%
\footnote{Because of the dominance of the ISR corrections this is not
  visible in the following figures.}%
, however, suppressed by the factor $Q_\Pt^2\alpha/(4\alphas/3)\sim
0.02$.  Although the relative QCD correction becomes rather huge close
to threshold, it should be realized that the total cross section
decreases rapidly there.
%The size of the large QCD corrections could also be reduced by
%imposing a cut on the maximal Higgs-boson energy that excludes small
%invariant masses of the $\Pt\bar\Pt$ system and thus the Coulomb-like
%singularity.  
In the region above threshold, where the cross section is largest, the
QCD correction is only of the order of a per cent.  For CM energies
and Higgs-boson masses far above threshold, the QCD corrections even
turn negative and reduce the cross section by about 5\% at
$\sqrt{s}=1\TeV$. This behaviour is expected from the effective
Higgs-boson approximation \cite{Dawson:1997im}, as also argued in
\citere{Dittmaier:1998dz}.

The photonic corrections are negative for CM energies below $1\TeV$
and not too small Higgs-boson masses, \ie not too far away from
threshold.  This is due to large virtual photonic corrections which
are not cancelled by the corresponding real radiation owing to the
decreasing phase-space volume for real-photon emission with increasing
$\MH$ or decreasing $\sqrt{s}$.  Therefore, the photonic corrections
are large and negative, in particular, at the relatively small
scattering energy of $\sqrt{s}=500\GeV$, where they reach $-25\%$ to
$-65\%$ for $\MH=100$--$150\GeV$.  In this situation resummation of
the large photonic corrections is mandatory. The bulk of these
contributions is included in our calculation (higher-order ISR) and
for $\sqrt{s}=500\GeV$ amounts to 3--20\%.  Away from threshold, the
photonic ${\cal O}(\alpha)$ corrections grow with increasing
$\sqrt{s}$ and reach $+7\%$ for $\MH=115\GeV$ and $\sqrt{s}=1.5\TeV$.
The higher-order ISR stays below 1\% for $\sqrt{s}$ at least $150\GeV$
above threshold.

The genuine weak corrections strongly depend on the scattering energy,
while the dependence on the Higgs-boson mass is moderate. For
$\sqrt{s}=500\GeV$ they are about 15--20\%, i.e.\ large and positive.
For increasing CM energy they decrease and are at the per-cent level
around $\sqrt{s}\sim800\GeV$.  For TeV energies they become more and
more negative and reach the order of $-10\%$ around $1.5\TeV$. While
the weak corrections are smaller than the QCD corrections in the
threshold region, both contributions are of similar size about
$250\GeV$ above threshold, \ie near the peak of the cross section. At
high energies the size of the weak corrections grows faster than the
one of the QCD corrections.  Such a behaviour is typical if Sudakov
logarithms like $\alpha\ln^2(s/\MW^2)$ dominate the weak corrections
at high energies.  Note that we use the $\GF$-scheme, and that the
weak corrections are shifted by $\sim\pm10\%$ when transformed to
other schemes like the $\alpha(0)$ or the $\alpha(s)$ schemes (see
also \citere{Denner:2003yg}).  The spikes at $\MH=2\MW,2\MZ$ result
from thresholds and are well-known from the process
$\Pep\Pem\to\PZ\PH$ \cite{Fleischer:1982af}.

Photonic and weak corrections partially cancel each other, and the
resulting electroweak corrections increase very weakly with $\sqrt{s}$
apart from the region close to threshold, where the electroweak
corrections decrease fast with decreasing $\sqrt{s}$ and reach of the
order of $-20\%$ at threshold.  Away from threshold, they are at the
level of $-5\%$, $-8\%$, and $-11\%$ for $\MH=115\GeV$, $150\GeV$, and
$200\GeV$, respectively, and thus become increasingly negative with
increasing Higgs-boson mass.

\section{Comparison with other calculations}
\label{se:comp}

The results on the QCD corrections given in \refta{ta:xsection} have
been reproduced with the (publically available) computer code based on
the calculation of \citere{Dittmaier:1998dz}. We found agreement
within the statistical integration errors.

For a comparison of the electroweak ${\cal O}(\alpha)$ corrections
with the results of \citere{You:2003zq} we changed our input
parameters to the ones quoted there and switched to the
$\alpha(0)$-scheme, where $\GF$ is ignored in the input and all
couplings are deduced from $\alpha(0)$.
In \refta{ta:Ren-You} we compare some representative numbers%
\footnote{These numbers were kindly provided to us by Zhang Ren-You
and You Yu quoting a statistical error below 1\%.}
from the calculation of \citere{You:2003zq} with the corresponding
results from our Monte Carlo generator. 
\begin{table}
\newcommand{\phm}{\phantom{-}}
$$\begin{array}{c@{\quad}c@{\quad}l@{\quad}l@{\quad}l@{\quad}l}
\hline
\sqrt{s}~[\mathrm{GeV}]& \MH~[\mathrm{GeV}]&
\ \sigma_{\mathrm{tree}}~[\mathrm{fb}] & 
\ \sigma_{{\cal O}(\alpha)}~[\mathrm{fb}]&
\ \delta_{\mathrm{ew}}~[\%]   & \\
\hline
 500 & 115 & 0.43343 & 0.4173 & -4.26   & 
\citere{You:2003zq} \\
     & 115 & 0.43341(6) & 0.4150(2) & -4.25(5)   & 
\mbox{this work} \\
\hline
 500 & 150 & 4.8142\cdot10^{-4} & 3.401\cdot10^{-4} & -29.35   & 
\citere{You:2003zq} \\
     & 150 & 4.8140(8) \cdot10^{-4} & 3.168(4)\cdot10^{-4} & -34.19(8)   & 
\mbox{this work} \\
\hline
 600 & 200 & 0.15359 & 0.1439 & -6.34   & 
\citere{You:2003zq} \\
     & 200 & 0.15359(2) & 0.14194(7) & -7.58(4)   & 
\mbox{this work} \\
\hline
 800 & 115 & 2.44 & 2.60 & \phm6.52   & 
\citere{You:2003zq} \\
     & 115 & 2.4434(3) & 2.5913(7) & \phm6.06(2)   & 
\mbox{this work} \\
\hline
 800 & 150 & 1.58 & 1.63 & \phm3.60   & 
\citere{You:2003zq} \\
     & 150 & 1.5749(2) & 1.6243(4) & \phm3.14(2)   & 
\mbox{this work} \\
\hline
 800 & 200 & 0.8341 & 0.8454 & \phm1.36   & 
\citere{You:2003zq} \\
     & 200 & 0.8340(1) & 0.8357(2) & \phm0.21(2)   & 
\mbox{this work} \\
\hline
1000 & 115 & 2.04 & 2.19 & \phm7.29   & 
\citere{You:2003zq} \\
     & 115 & 2.0443(3) & 2.1935(5) & \phm7.30(2)   & 
\mbox{this work} \\
\hline
1000 & 150 & 1.47 & 1.53 & \phm4.47   & 
\citere{You:2003zq} \\
     & 150 & 1.4664(2) & 1.5273(4) & \phm4.15(2)   & 
\mbox{this work} \\
\hline
1000 & 200 & 0.9372 & 0.9567 & \phm2.09   & 
\citere{You:2003zq} \\
     & 200 & 0.9370(1) & 0.9492(2) & \phm1.29(2)   & 
\mbox{this work} \\
\hline
2000 & 115 & 0.7614 & 0.7919 & \phm4.02   & 
\citere{You:2003zq} \\
     & 115 & 0.7613(1) & 0.8214(4) & \phm7.90(5)   & 
\mbox{this work} \\
\hline
2000 & 150 & 0.6270 & 0.6297 & \phm0.43   & 
\citere{You:2003zq} \\
     & 150 & 0.6269(1) & 0.6526(3) & \phm4.11(5)   & 
\mbox{this work} \\
\hline
2000 & 200 & 0.4968 & 0.4790 & -3.55   & 
\citere{You:2003zq} \\
     & 200 & 0.49659(8) & 0.5003(3) & \phm0.74(5)   & 
\mbox{this work} \\
\hline
\end{array}$$
\caption{Total cross section in lowest order and including the full
  electroweak $\Oa$ corrections as well as the relative corrections for
  various CM energies and Higgs-boson masses for the input-parameter
  scheme of \citere{You:2003zq}. The statistical errors of
  \citere{You:2003zq} are estimated  by the authors to be below 1\%
  (\cf footnote \thefootnote). }
\label{ta:Ren-You}
\end{table}
The numbers in parentheses give the errors in the last digits of our
calculation.  The tree-level cross sections coincide within 0.03\%.
Most of the numbers for the one-loop corrected cross sections agree
within 1--2\%, \ie roughly within the estimated error of
\citere{You:2003zq}.  However, for the corrected cross sections at
$\sqrt{s}=2\TeV$, \ie at high energies, and the one very close to
threshold, \ie for $\sqrt{s}=500\GeV$ and $\MH=150\GeV$, we find
differences of 4\% and 7\%, respectively.  The same holds for the
relative corrections. Ours are larger by about 4\% at $\sqrt{s}=2\TeV$
and smaller by about 5\% for the selected cross section close to
threshold.

Finally, we have also compared the electroweak ${\cal O}(\alpha)$
corrections with \citere{Belanger:2003nm}, where also the
$\alpha(0)$-scheme has been used. In \refta{ta:Belanger} we list the
results of Table~2 of \citere{Belanger:2003nm} and the separate
relative photonic and weak corrections of Table~3 of that paper
together with the corresponding results from our Monte Carlo
generator. Again the numbers in parentheses give the errors in the
last digits. We reproduce the results for the lowest-order cross
section within the integration errors, which are about
2--$3\times10^{-4}$.  The results for the cross section including
electroweak corrections coincide to better than 0.1\% which is of the
order of the integration error of the results of
\citere{Belanger:2003nm}.  The relative electroweak, weak, and
photonic corrections agree also within 0.1\%. This holds as well for
the QCD corrections (not shown in \refta{ta:Belanger}).
%For large Higgs-boson masses the relative photonic corrections of
%B\'elanger et al. are systematically about 0.1\% larger than ours.
\begin{table}
\newcommand{\phm}{\phantom{-}}
$$\begin{array}{c@{\quad}c@{\quad}l@{\quad}l@{\quad}l@{\quad}l@{\quad}l@{\quad}l}
\hline
%\rule[-1mm]{0mm}{5mm} 
%\sqrt{s}~[\mathrm{GeV}]& \MH~[\mathrm{GeV}]&
%\sigma_{\mathrm{tree}}~[\mathrm{fb}] & \sigma_{{\cal O}(\alpha)}~[\mathrm{fb}]&
%\delta_{\mathrm{ew}}~[\%]  & \delta_{\mathrm{weak}}~[\%]  & \delta_{\mathrm{phot}}~[\%]  & \\
\sqrt{s} & \MH &
\sigma_{\mathrm{tree}}~[\mathrm{fb}] & \sigma_{{\cal O}(\alpha)}~[\mathrm{fb}]&
~~\delta_{\mathrm{ew}}~[\%]  & \delta_{\mathrm{weak}}~[\%]  & 
~\:\delta_{\mathrm{phot}}~[\%]  & \\{}
[\mathrm{GeV}] &[\mathrm{GeV}]& & & & & & \\
\hline
%\hline \rule[-1mm]{0mm}{5mm}
 600 & 120 & 1.7293(3) & 1.738(2) & \phm0.5   & 16.5    & -16.0 & 
\citere{Belanger:2003nm} \\
%     & 120 & 1.7288(2) & 1.7365(10) & \phm0.45(6) & 16.47(6) & -16.02(1) & 
%\mbox{this work} \\  % mar
     & 120 & 1.7292(2) & 1.7368(6) & \phm0.44(3) & 16.49(3) & -16.03(1) & 
\mbox{this work} \\  % maw
\hline
%\cline{2-8}\\
%\rule[-1mm]{0mm}{5mm} 
600  & 180 & 0.33714(4) & 0.3126(3) & -7.3 & 18.4 & -25.7 & 
\citere{Belanger:2003nm}\\
\rule[-1mm]{0mm}{5mm} 
%~    & 180 & 0.33705(3) & 0.3123(2) & -7.36(3) & 18.34(6) & -25.70(1) &
% \mbox{this work} \\ %mar
~    & 180 & 0.33714(5) & 0.3124(1) & -7.34(3) & 18.38(3) & -25.72(1) &
 \mbox{this work} \\ %maw
\hline
 800 & 120 & 2.2724(5) & 2.362(4) & \phm3.9 & 9.5 & -5.6 &
\citere{Belanger:2003nm}\\
%     & 120 & 2.2717(4) & 2.3590(10) & \phm3.84(4) & 9.53(4) & -5.69(1) &
% \mbox{this work} \\%mar
     & 120 & 2.2723(3) & 2.3599(6) & \phm3.86(2) & 9.56(2) & -5.70(1) &
 \mbox{this work} \\%maw
\hline
 800 & 180 & 1.0672(3) & 1.050(2) & -1.6 & 9.1 & -10.7 &
\citere{Belanger:2003nm}\\
\rule[-1mm]{0mm}{5mm} 
%~    & 180 & 1.0665(1) & 1.0490(4) & -1.64(3) & 9.16(3) & -10.80(1) &
% \mbox{this work} \\%mar
~    & 180 & 1.0668(2) & 1.0494(2) & -1.63(2) & 9.18(2) & -10.81(1) &
 \mbox{this work} \\%maw
\hline 
1000 & 120 & 1.9273(5) & 2.027(4) & \phm5.2     & 5.8     &  -0.6  &
\citere{Belanger:2003nm}\\
%     & 120 & 1.9266(4) & 2.0249(6) & \phm5.10(2) & 5.80(2) &  -0.70(1) &
% \mbox{this work} \\%mar
     & 120 & 1.9271(3) & 2.0252(5) & \phm5.09(2) & 5.78(2) &  -0.70(1) &
 \mbox{this work} \\%maw
\hline
1000 & 180 & 1.1040(3) & 1.098(2) & -0.5     & 4.4     &  -4.9  &
\citere{Belanger:2003nm}\\
% ~   & 180 & 1.1036(2) & 1.0969(3) & -0.61(2) & 4.38(2) &  -4.98(1) &
%\mbox{this work} \\%mar
 ~   & 180 & 1.1039(2) & 1.0972(3) & -0.61(2) & 4.41(2) &  -5.00(1) &
\mbox{this work} \\ %maw
\hline
\end{array}$$
\caption{Total cross section in lowest order and including the full
  electroweak $\Oa$ corrections as well as the relative electroweak,
  weak, and photonic corrections for various CM energies and
  Higgs-boson masses for the input-parameter scheme of
  \citere{Belanger:2003nm}} 
\label{ta:Belanger}
\end{table}

\section{Summary}
\label{se:sum}
We have presented results from a calculation of electroweak radiative
corrections to the process $\eetth$, which is important for a precise
determination of the top--Higgs Yukawa coupling.  In detail, we have
discussed the impact of photonic corrections at and beyond ${\cal
  O}(\alpha)$, the genuine weak ${\cal O}(\alpha)$ corrections, and
the ${\cal O}(\alphas)$ QCD corrections.

The photonic and weak corrections both reach the order of $\sim 10\%$
and show characteristic dependences on the Higgs-boson mass and on the
scattering energy. Owing to a phase-space effect the (negative)
photonic corrections reduce the cross section more and more when the
threshold is approached, \ie with increasing Higgs-boson masses.
Close to threshold the resummation of the higher-order ISR corrections
becomes mandatory.  The weak corrections, which depend on the
Higgs-boson masses only moderately, range from about $+15\%$ at a CM
energy of $500\GeV$ to about $-10\%$ at $1.5\TeV$.  There are large
cancellations between photonic and weak corrections, and the final
size of the corrections depends strongly on the input-parameter
scheme.  These results clearly demonstrate the necessity to include
the electroweak corrections in predictions adequate for a future
high-luminosity $\Pep\Pem$ collider.

We have compared our results with those of other groups. The QCD
corrections have been successfully checked against previous
calculations. The electroweak corrections agree well with the results
of the recent calculation \cite{Belanger:2003nm} but are at variance
with the results of \citere{You:2003zq} at high energies and close to
threshold.

\section*{Acknowledgement}

We thank Zhang Ren-You and You Yu for sending us some representative
numbers for comparison with the calculation of \citere{You:2003zq}.
This work was supported in part by the Swiss Bundesamt f\"ur Bildung
und Wissenschaft and by the European Union under contract
HPRN-CT-2000-00149.

\end{document}